\documentclass[paper]{JHEP}
\usepackage{epsfig}
\def\as{\alpha_s}
\def\mI{\mu_{\mbox{\tiny I}}}
\def\re#1{(\ref{#1})}
\def\half{{\textstyle {1\over2}}}
\def\beq{\begin{equation}}   \def\eeq{\end{equation}}
\def\beeq{\begin{eqnarray}}   \def\eeeq{\end{eqnarray}}
\title{QCD Power Corrections from a Simple Model for the Running Coupling}
\author{Bryan R.\ Webber\thanks{Research
supported in part by the U.K. Particle Physics and Astronomy Research
Council and by the EC Programme ``Training and Mobility of Researchers",
Network ``Hadronic Physics with High Energy Electromagnetic Probes",
contract ERB FMRX-CT96-0008.}\\
        Cavendish Laboratory, University of Cambridge,\\
        Madingley Road, Cambridge CB3 0HE, U.K.\\
        E-mail: \email{webber@hep.phy.cam.ac.uk}, etc.}
\abstract{A simple parametrization of the QCD running coupling at
low scales is introduced and used to illustrate various schemes for
the estimation of non-perturbative power corrections. The
`infrared matching' scheme proposed earlier gives satisfactory
results when combined with next-to-leading (or higher) order
perturbative predictions. Estimates based on renormalons are shown
to be inconsistent with universal behaviour of the running coupling.}
\preprint{Cavendish--HEP--98/08\\hep-ph/9805484}
\begin{document}
\section{Introduction}
In several recent papers \cite{BPY}-\cite{Haut} the hypothesis of
a universal low-energy behaviour of the QCD running coupling has been
used to estimate power-behaved corrections to various QCD observables.
While the only essential parameters are a few moments of the coupling
in the infrared region, the absence of a satisfactory explicit
model for the non-perturbative contributions has made the arguments
hard to follow and the approximation scheme unclear.  In the present
paper, a simple model is introduced with sufficiently good
properties to allow quantitative comparisons of various proposed
schemes for the estimation of power corrections. 

Let us start by recalling that the running coupling has the general form
\beq\label{ask2}
\as(k^2) \equiv \frac{1}{b_0}a(k^2/\Lambda^2)
\eeq
where for $N_f$ active flavours
\beq
b_0 = \frac{33-2N_f}{12\pi}
\eeq
and to 1-loop order at high energies
\beq\label{as1l}
a(x) \sim \frac{1}{\ln x}\;.
\eeq
A fundamental problem with simply continuing this expression
to low energies is the Landau pole at $k^2=\Lambda^2$
($x=1$). On general causality grounds, at least if the running
coupling is to be related analytically to some physical observable,
we expect singularities of the function $\as(k^2)$ to occur
on the negative real axis only.

A number of explicit models have been proposed
\cite{DoKhTr95}-\cite{AleArb} for including non-perturbative
contributions at low $k^2$ and cancelling the Landau pole.
A particularly simple one with many good features is
the `analytic model' \cite{ShiSol,Grun},
\beq\label{ashisol}
a(x) = \frac{1}{\ln x}+\frac{1}{1-x}\;.
\eeq
This expression has no Landau pole,
only a branch point at $k^2=0$. However, the extra term introduces
a $1/k^2$ correction to the logarithmic running of the coupling
at large $k^2$. This would have to be cancelled somehow in quantities
that are proportional to $\as(Q^2)$ in leading order but are not
expected to have $1/Q^2$ corrections, such as the $e^+e^-$ total
cross section.  In addition the numerical value of $\as$ becomes
rather large, $\as(0) = 1/b_0 = 1.4$ for $N_f=3$, which
calls into question the notion of an expansion in powers of
$\as$ at low scales. One might also prefer to have an expression
with free parameters, other than the overall momentum scale $\Lambda$,
that could be fitted to experimental data. The expressions suggested
in Ref.~\cite{DoKhTr95} do have free parameters,
but they also have non-analytic terms, unwanted power corrections
and/or singularities at unphysical values of $k^2$.

\section{Model for the running coupling}
If we require
\begin{itemize}
\item No power corrections larger than $1/k^{2p}$,
\item Singularities on the negative real $k^2$ axis only,
\item Some freedom to adjust the form and value at low $k^2$,
\end{itemize}
then the following generalization of Eq.~\re{ashisol} suggests itself:
\beq\label{abw}
a(x) = \frac{1}{\ln x}+\frac{x+b}{(1-x)(1+b)}
\left(\frac{1+c}{x+c}\right)^p\;.
\eeq
For definiteness we shall consider here $b=1/4$, $c=4$, $p=4$, i.e.
\beq\label{adef}
a(x) = \frac{1}{\ln x}+125\frac{1+4x}{(1-x)(4+x)^4}\;,
\eeq
which leads to the form of $\as$ shown in Fig.~1, where the values
$N_f=3$, $\Lambda=0.25$ GeV have been chosen, giving $\as(Q^2)=0.118$
at $Q=91$ GeV. For this choice of parameters,  $\as(k^2)$ lies very
close to the perturbative expression (dot-dashed)
above $k=1$ GeV, the difference falling like $1/k^8$.
Below 1 GeV, however, it remains finite and positive,
reaching a maximum value of 0.8 at $k=0.4$ GeV. As discussed
below, the moments of this function are consistent with
those deduced from data on power corrections. The multipole
on the negative real axis looks somewhat unphysical,
but should be regarded only as an approximation to
a more complicated singularity structure.

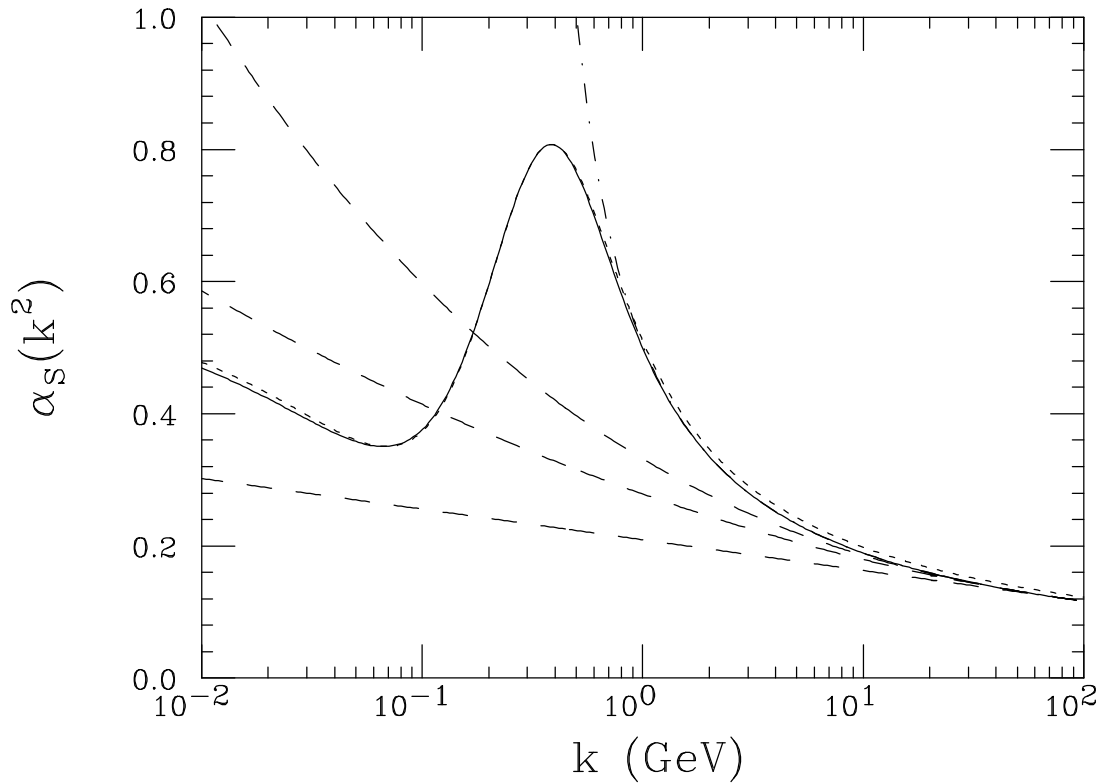
\EPSFIGURE{alf.eps}{Model for the running coupling (solid curve),
compared with one-loop perturbative form (dot-dashed) and
expansions to first, second and third order in $\as(Q^2)$
(dashed, $Q=91$ GeV). The dotted curve shows the 2-loop
modification \re{2loop}.}

In order to relate the parameter $\Lambda$ in Eq.~\re{ask2} to
the scale parameter in some particular renormalization scheme,
Eq.~\re{abw} should be generalized to reproduce the 2-loop
behaviour of $\as(k^2)$ at large $k^2$. As has been discussed
for the simpler form \re{ashisol} \cite{ShiSol}, this can be
achieved without spoiling the analytic properties through
a replacement of the form
\beq\label{2loop}
x \to x\left[1+\left(\frac{\ln x}{2\pi}\right)^2\right]^b
\eeq
where
\beq\label{bdef}
b=\frac{b_1}{2b_0^2} =\frac{3(153-19N_f)}{(33-2N_f)^2}\;.
\eeq
As shown by the dotted curve in Fig.~1, this makes only a very
small difference at fixed values of the parameters, although
of course these values would also need some readjustment.
For simplicity we shall therefore use the 1-loop formula \re{adef}
in the remainder of this paper.

\section{Power corrections}
As a model for a QCD observable which is of perturbative
order $\as^q$ and is expected to have a leading power correction
of order $1/Q^p$, consider the integral \cite{DokUra}
\beq\label{Fpq}
F_{p,q}(Q) =\frac{p}{Q^p}\int_0^Q \frac{dk}{k}k^p\left[\as(k^2)\right]^q\;.
\eeq
Then to lowest order in $\as(Q^2)$ we have $F_{p,q}=F^{(0)}_{p,q}$ where
\beq\label{F0pq}
F^{(0)}_{p,q}(Q) = \left[\as(Q^2)\right]^q\;.
\eeq
Using the explicit expression given by Eqs.~\re{ask2} and \re{adef},
we can evaluate the integral exactly, as shown for $p=q=1$ by the
solid curve in Fig.~2.  The difference between $F_{p,q}$ and
$F^{(0)}_{p,q}$ arises from the running of $\as(k^2)$ in the
integrand, which receives both perturbative and non-perturbative
contributions.  It is not possible to disentangle these
unambiguously, because the two terms in Eq.~\re{adef} separately
give divergent contributions. A perturbative expansion in powers
of $\as(Q^2)$ sees only the first term; expanding to
${\cal O}(\as^{q+N})$ means using
\beq\label{askq}
\left[\as(k^2)\right]_N^q \equiv \left[\as(Q^2)\right]^q
\sum_{n=0}^N \frac{(n+q-1)!}{n!\,(q-1)!}\left[b_0\as(Q^2)
\ln\left(Q^2/k^2\right)\right]^n
\eeq
in Eq.~\re{Fpq} to obtain
\beq\label{Fpqpert}
F^{(N)}_{p,q}(Q) = \left[\as(Q^2)\right]^q\sum_{n=0}^N
\frac{(n+q-1)!}{(q-1)!}\left[\frac{2b_0}{p}\as(Q^2)\right]^n\;,
\eeq
which diverges as $N\to\infty$.  This is the infrared
renormalon problem \cite{DokUra}-\cite{AkZak}.
As one would expect from the corresponding
curves in Fig.~1, the perturbative estimate \re{Fpqpert} of
$F_{p,q}$ undershoots for low values of $N$, but becomes
arbitrarily large for sufficiently high $N$ (Fig.~2).

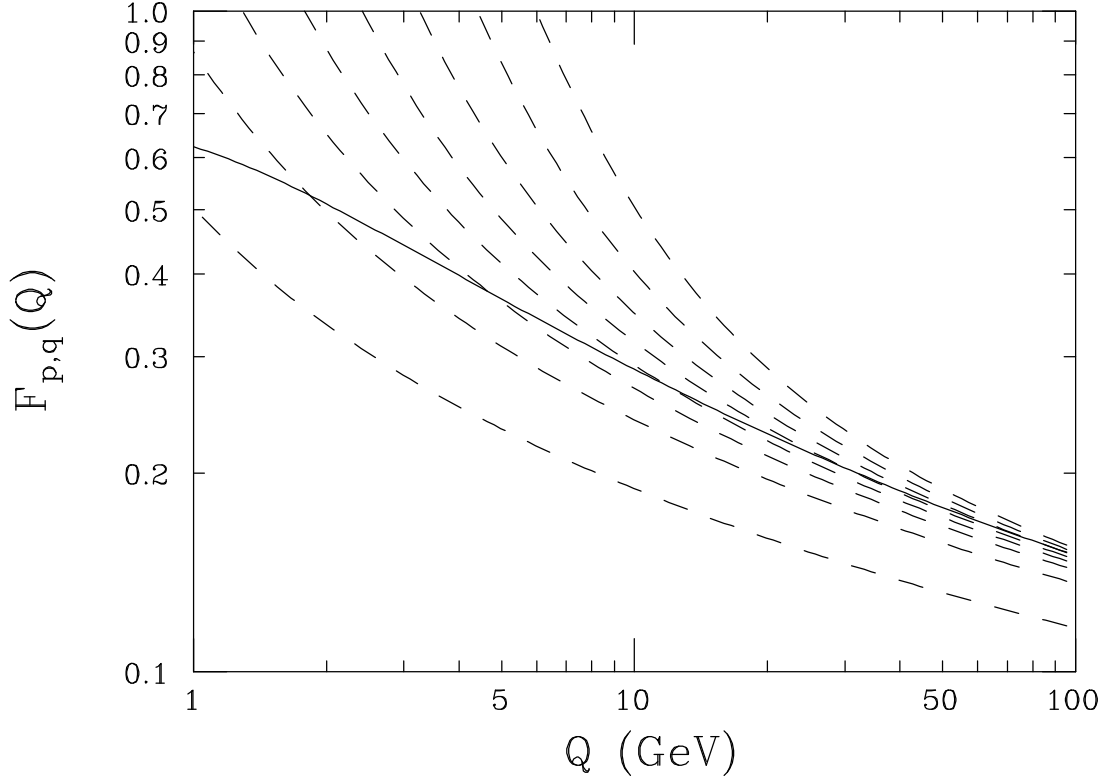
\EPSFIGURE{fpqn.eps}{Exact value of the integral \re{Fpq}
for $p=q=1$ (solid curve), compared with perturbative
estimates \re{Fpqpert} for $N=0,\ldots,7$ (dashed).}

To correct the ${\cal O}(\as^{q+N})$ perturbative estimate
\re{Fpqpert} we clearly have to write
\beq\label{Fpqcorr}
F_{p,q}(Q) = F^{(N)}_{p,q}(Q)+ \delta F^{(N)}_{p,q}(Q)
\eeq
where the correction term is
\beq\label{delFpq}
\delta F^{(N)}_{p,q}(Q) =
\frac{p}{Q^p}\int_0^Q\frac{dk}{k}k^p\left(\left[\as(k^2)\right]^q
-\left[\as(k^2)\right]_N^q\right)\;.
\eeq
In the approximation scheme proposed in Ref.~\cite{DokWeb},
one notes that the
two terms in the integrand of Eq.~\re{delFpq} must become similar
at values of $k$ in the perturbative region, say $k>\mI$, and
that this region cannot contribute to the renormalon divergence.
Therefore we neglect the integrand above $\mI$,
the `infrared matching' scale, and write
\beeq\label{delFpqsim}
\delta F^{(N)}_{p,q}(Q) &\simeq&
\frac{p}{Q^p}\int_0^{\mI} dk\, k^{p-1}\left(\left[\as(k^2)\right]^q
-\left[\as(k^2)\right]_N^q\right) \nonumber \\
&=&  \left(\frac{\mI}{Q}\right)^p F_{p,q}(\mI)- G^{(N)}_{p,q}(\mI,Q)
\eeeq
where
\beq\label{Gpqpert}
G^{(N)}_{p,q}(\mI,Q) = \left[\as(Q^2)\right]^q\sum_{n=0}^N
\frac{(n+q-1)!}{n!\,(q-1)!}\left[\frac{2b_0}{p}\as(Q^2)\right]^n
\Gamma(n+1,p\ln(Q/\mI))\;,
\eeq
$\Gamma$ being the incomplete gamma function,
\beq\label{Gamma}
\Gamma(n+1,z) = \int_z^\infty t^n\,e^{-t}\,dt\;.
\eeq
The point of this exercise is that all the necessary non-perturbative
information is now contained in the $Q$- and $N$-independent
parameters $F_{p,q}(\mI)$, which represent weighted averages of
$\as^q$ over the infrared region $0<k<\mI$. The resulting contribution
to $F_{p,q}(Q)$ is a power correction of order $1/Q^p$. The second term
on the right-hand side of Eq.~\re{delFpqsim} subtracts the divergent
renormalon part of the perturbative prediction.  After this
subtraction, the total correction $\delta F^{(N)}_{p,q}(Q)$
is not simply power-behaved, as may be seen from the difference
between the exact result and any of the fixed-order curves in
Fig.~2.

\TABLE{
\begin{tabular}{|c|r r r r|}\hline
$q$ &  $p=1$ &  $p=2$ &  $p=3$ &  $p=4$ \\ \hline
 1  &  0.511 &  0.450 &  0.410 &  0.388 \\  
 2  &  0.283 &  0.218 &  0.176 &  0.155 \\
 3  &  0.170 &  0.114 &  0.081 &  0.064 \\ \hline
\end{tabular}
\caption{Values of $F_{p,q}(2\;\mbox{GeV})$.}
}

For the model of the running coupling shown in Fig.~1, we see that
$\mI=2$ GeV is a reasonable matching scale, for which one finds
the values of $F_{p,q}(\mI)$ given in Table~1. The results
$F_{1,1}(2\;\mbox{GeV})\simeq F_{2,1}(2\;\mbox{GeV})\simeq 0.5$ are
consistent with the values
deduced \cite{BPY,DokWeb,DasWeb,DLMS,Web94}
from experimental data on event shapes \cite{DELPHI,JADEOPAL,H1}
and structure functions \cite{NMC,CCFR},
respectively.\footnote{Note that in Refs.~\cite{DokWeb,DasWeb}
the notation $F_{p,1}(\mI) = \bar\alpha_{p-1}(\mI)$ was used.}

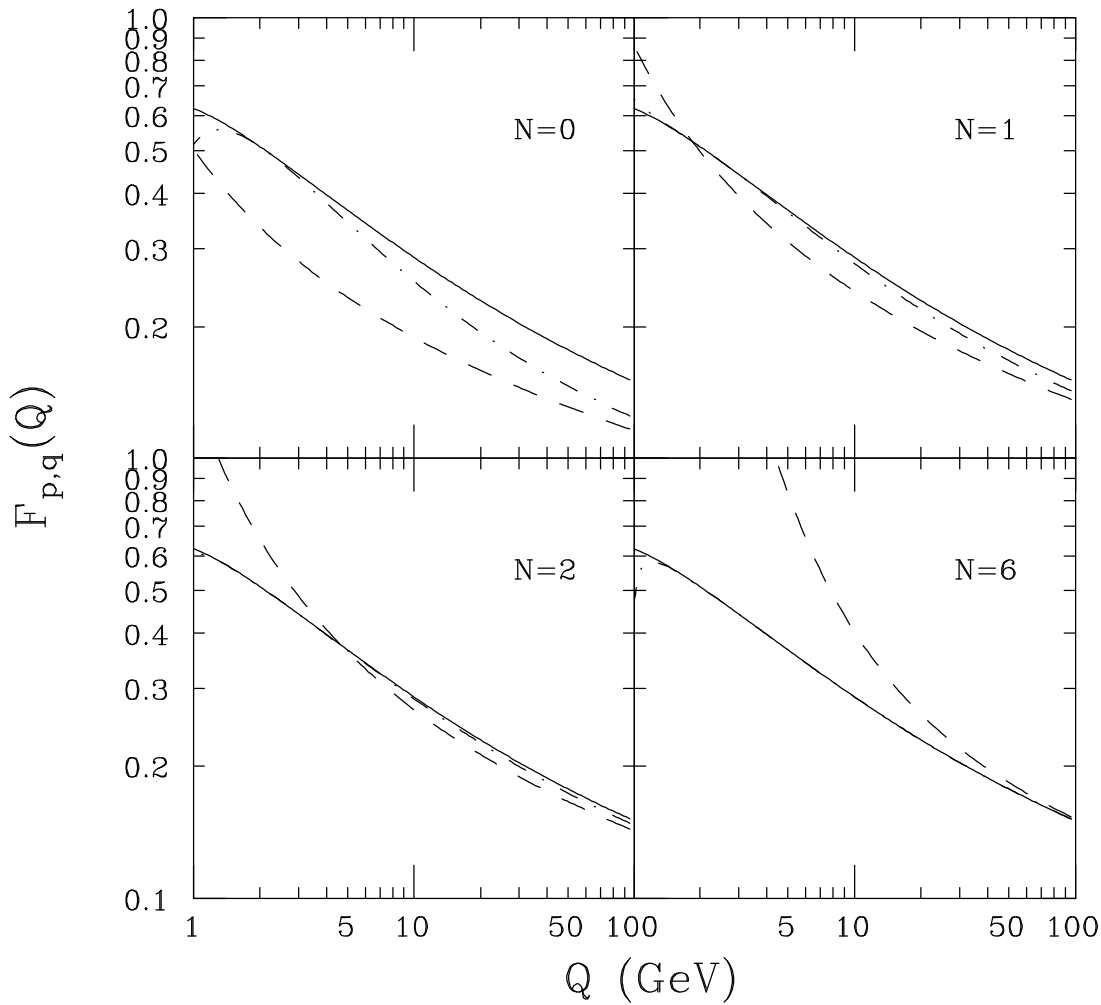
\EPSFIGURE{fpqmatch.eps}{Exact value of the integral \re{Fpq}
for $p=q=1$ (solid curves), compared with perturbative (dashed)
and infrared matching (dot-dashed) estimates for $N=0,1,2,6$.}

Using these values in Eq.~\re{delFpqsim}, we obtain the 
($\mI=2$ GeV) `matching estimates' of the integral \re{Fpq},
shown by the dot-dashed
curves in Fig.~3 for $N=0$, 1, 2, 6. The $N=1$ case corresponds to
the most common situation, where we have to match to a NLO perturbative
calculation. There is still some discrepancy at large $Q$ because
the matching in the interval $\mI<k<Q$ is not perfect.
The discrepancy is reduced when the NNLO term is added ($N=2$).
The  $N=6$ case is shown to illustrate the cancellation
of the renormalon and the improvement in the matching
when very many perturbative terms are included.

Since we know the exact form of the integrand, we could of
course improve the matching estimates at large $Q$ by choosing
a larger value of the matching scale $\mI$. However in reality
we usually know only the behaviour as $k^2/Q^2\to 0$, and
$\mI$ must be kept small in order for this behaviour to
predominate in the correction term \re{delFpq}.

\section{Renormalons}
A common way to deal with the renormalon divergence of the perturbative
estimate \re{Fpqpert} as $N\to\infty$ is to Borel transform and define
the sum by the principal value of the Borel
integral \cite{Zakh,AkZak,BBNNA,Neu,NaSey,Stein}. Recall that if
\beq\label{Fas}
F(\as) = \sum_{n=0}^\infty c_n\as^{n+1}
\eeq
then the Borel transform is
\beq\label{tFas}
\tilde F(u) = \sum_{n=0}^\infty \frac{c_n}{n!}u^n
\eeq
and the sum would be defined as
\beq\label{Fasint}
F(\as) = P \int_0^\infty du\,\tilde F(u)\,e^{-u/\as}
\eeq
where $P$ indicates the principal value. In the case of
Eq.~\re{Fpqpert} we have
\beq\label{tFpqu}
\tilde F^{(\infty)}_{p,q}(u) = \frac{u^{q-1}}{(q-1)!}\;
\frac{p}{p-2b_0u}
\eeq
and hence the `renormalon estimate' of $F_{p,q}(Q)$ is
\beq\label{Fpqren}
F^{(ren)}_{p,q}(Q) = \left[\as(Q^2)\right]^q
H_q\left(\frac{p}{2b_0\as(Q^2)}\right)
\eeq
where
\beq\label{Hq}
H_q(z) = \frac{z^q}{(q-1)!} \left[e^{-z}\mbox{Ei}(z)
-\sum_{n=0}^{q-2} n!\,z^{-n-1}\right]
\eeq
and Ei represents the exponential integral function,
\beq\label{eidef}
\mbox{Ei}(z) = -P\int_{-z}^\infty \frac{e^{-t}}{t}\,dt\;.
\eeq

The uncertainty in the estimate \re{Fpqren} is normally taken
to be given by the residue of the renormalon pole at $u=p/2b_0$:
\beq\label{dFpqren}
\delta F^{(ren)}_{p,q}(Q) = \left[\as(Q^2)\right]^q
\delta H_q\left( \frac{p}{2b_0\as(Q^2)}\right)
\eeq
where
\beq\label{dHq}
\delta H_q(z) = \frac{z^q}{(q-1)!}\,e^{-z}\;.
\eeq

For $q=1$, the renormalon estimate \re{Fpqren} corresponds
directly to a principal-value definition of the integral
\re{Fpq}, i.e.\ to using the perturbative form \re{as1l}
for $\as(k^2)$ everywhere apart from smoothing the Landau pole
over a small region around $k=\Lambda$. This means
that $\as(k^2)<0$ for $k<\Lambda$, leading to a slight
underestimation of $F_{p,q}$ when $p=q=1$ (Fig.~4). As $p$
increases, the negative contribution from  $k<\Lambda$ is
suppressed and the region of positive $\as$ dominates, leading
to a slight overestimate. Nevertheless, for the simple model
\re{adef} assumed here for $\as$, the renormalon estimate is
quite satisfactory when $q=1$, within the assumed uncertainty
\re{dFpqren}, shown by the error bars.

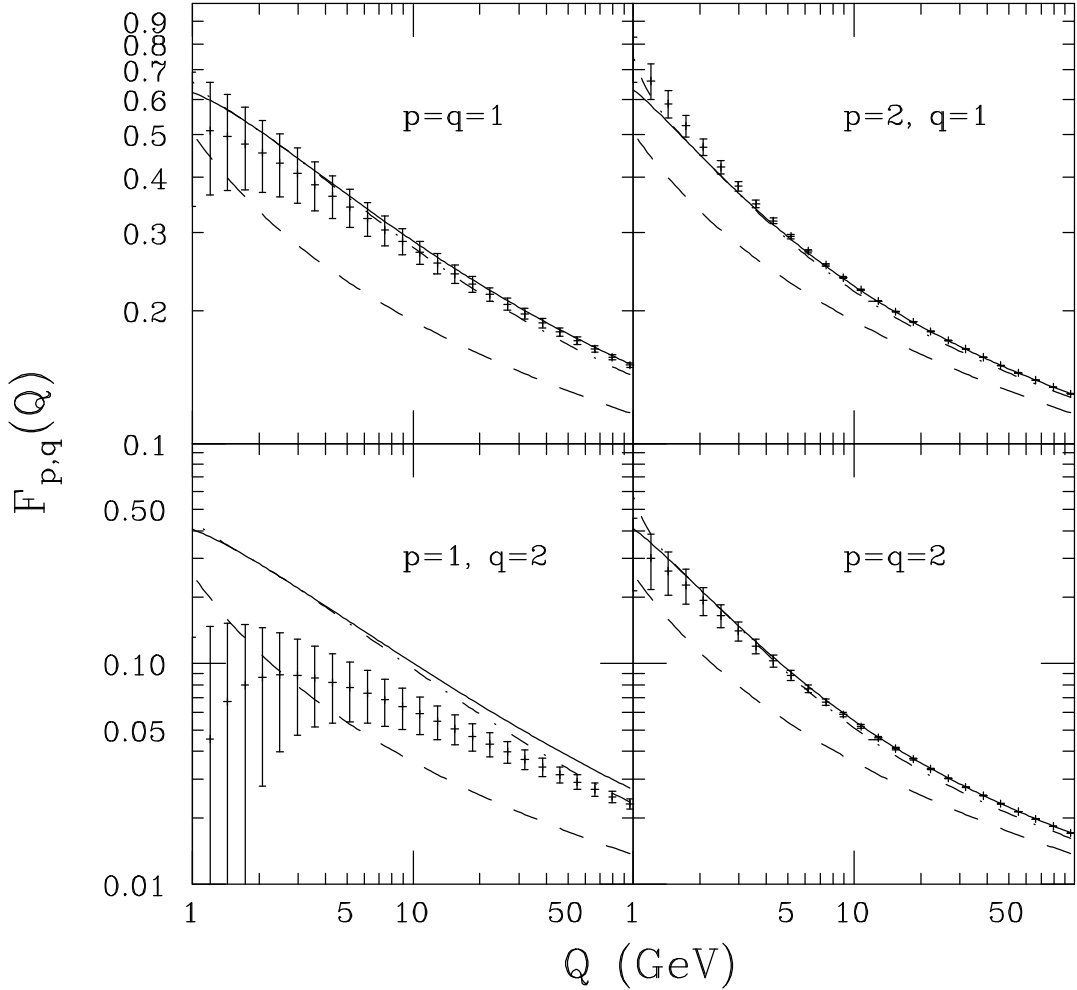
\EPSFIGURE{fpqren.eps}{Exact values of the integral \re{Fpq}
for (solid curves), and renormalon estimates (points). For
comparison the LO perturbative (dashed) and NLO matched
(dot-dashed) estimates are also shown.}

For higher orders, $q>1$, the situation is different.
If we use the perturbative expression for $\as(k^2)$,
the integrand in \re{Fpq} has a multiple pole and the
meaning of the renormalon estimate \re{Fpqren} is not obvious.
A useful way to interpret it (valid also for $q=1$) is as
follows.  The function $F(Q) = F_{p,q}(Q)$ satisfies the
differential equation
\beq\label{dFdQ}
\frac{Q}{p}\frac{dF}{dQ} + F = \left[\as(k^2)\right]^q\;,
\eeq
which has the general solution
\beq\label{FQ0}
F(Q) =\frac{p}{Q^p}\int_{Q_0}^Q \frac{dk}{k}k^p\left[\as(k^2)\right]^q
\eeq
where $Q_0$ is an arbitrary constant of integration.
The renormalon estimate \re{Fpqren} corresponds to a particular
choice of $Q_0>0$, namely the solution of
\beq\label{asQ0}
\as(Q_0^2) = \frac{p}{2b_0 z_q}\;,
\eeq
$z=z_q$ being the point where $H_q(z)=0$, i.e.\ where
\beq
\mbox{Ei}(z)=e^z\sum_{n=0}^{q-2} n!\,z^{-n-1}\;.
\eeq
Explicit values are
\beq\label{zqs}
z_1=0.373\,,\>\>z_2=1.347\,,\>\>z_3=2.342\,,\>\>z_q\sim q-0.67\;.
\eeq

\TABLE{
\begin{tabular}{|c|r r r r|}\hline
$q$ &  $p=1$ &  $p=2$ &  $p=3$ &  $p=4$ \\ \hline
 1  &  1.45  &  1.20  &  1.13  &  1.10  \\  
 2  &  3.85  &  1.96  &  1.57  &  1.40  \\
 3  & 10.40  &  3.22  &  2.18  &  1.80  \\ \hline
\end{tabular}
\caption{Values of $Q_0(p,q)/\Lambda$.}
}

Clearly the solution \re{FQ0} corresponds (for $Q>Q_0$) to setting
$\as(k^2)\equiv 0$ throughout $0<k<Q_0$. Thus the renormalon
estimate \re{Fpqren} is equivalent to using the perturbative
expression for $\as(k^2)$ everywhere above an `effective cutoff'
$Q_0(p,q)$ given by Eq.~\re{asQ0}, and setting $\as(k^2)\equiv 0$
below that value. Using the 1-loop expression for $\as(Q_0^2)$,
we obtain
\beq
Q_0(p,q) =\Lambda\,e^{z_q/p}\sim \Lambda\,e^{(q-0.67)/p}\;.
\eeq
Note that the effective cutoff $Q_0(p,q)$ grows rapidly
with increasing $q$, especially for $p=1$ (Table~2).

When $q=1$, setting $\as(k^2)=0$ for $k<Q_0(p,1)$ is equivalent
to the principal value definition of the integral in \re{Fpq},
because the principal value of the integral from 0 to $Q_0(p,1)$
is zero.  For $q>1$, the higher effective cutoff
leads to underestimation of the integral (Fig.~4).
The fact that the cutoff on $\as$ is imposed at a scale
that depends on $p$ and $q$ makes the renormalon
approach look unnatural. Given that the analyticity
properties are destroyed by the principal value prescription,
one might as well choose to set $\as(k^2)=0$ below some fixed
value of $Q_0$, to be treated as a free parameter.

Another way of defining the divergent sum \re{Fpqpert} is by
truncating the series at its smallest term \cite{Zakh},
i.e.\ summing up to the integer nearest to
\beq
N(Q) = \frac{p}{2b_0\as(Q^2)}-q+\half
\eeq
and assigning an uncertainty equal to the smallest term.
This is valid if the series is an asymptotic expansion.
The resulting `optimal truncation' estimate is again a
solution of the differential equation \re{dFdQ}, but the
effective cutoff $Q_0(p,q)$ is now given by $N(Q_0)=0$, i.e.
\beq\label{asQ0tr}
\as(Q_0^2) = \frac{p}{2b_0(q-\half)}
\eeq
which is just Eq.~\re{asQ0} with $z_q$ replaced by $q-\half$.
In view of Eqs.~\re{zqs}, one then expects results very
similar to those of the renormalon approach, which
is indeed the case (Fig.~5).

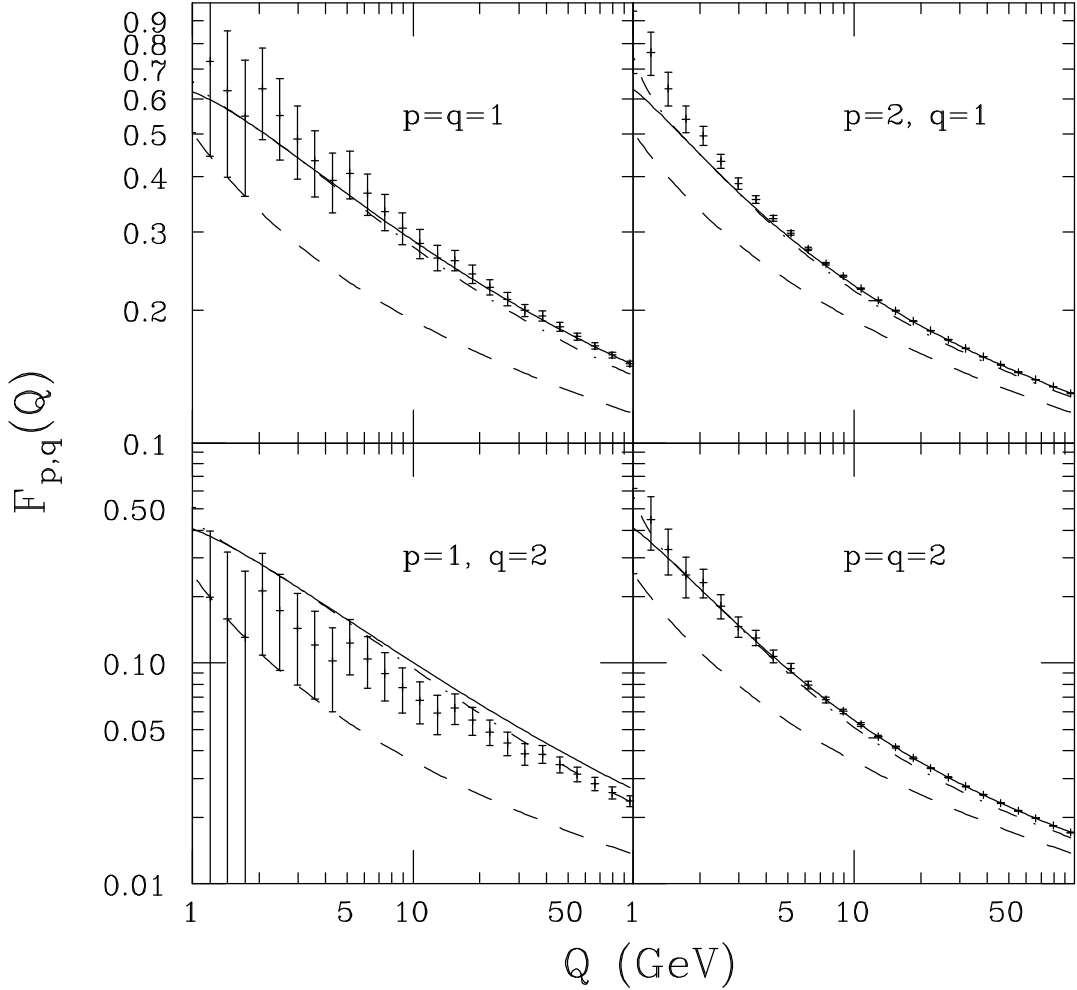
\EPSFIGURE{fpqtrun.eps}{Exact values of the integral \re{Fpq}
(solid curves), and optimally truncated estimates (points).
For comparison the LO perturbative (dashed) and NLO matched (dot-dashed)
estimates are also shown.}

\section{Conclusions}
The simple model of the running coupling introduced in Sect.~2
has allowed us to study various aspects of the estimation of QCD
power corrections in some detail. For this purpose we have used
the toy model observable $F_{p,q}(Q)$ defined in Eq.~\re{Fpq}.

The `infrared matching'
procedure introduced in Ref.~\cite{DokWeb} gives fairly
good results when matched with the NLO ($N=1$) perturbative
estimate, improving rapidly as higher orders are included (Fig.~3).
In principle it eliminates the renormalon problem completely,
at the price of introducing new non-perturbative parameters,
which are the moments of $\as$ at low scales (Table 1).
The simple model expresses these moments in terms of its
own non-perturbative parameters, $b$, $c$ and $p$ in Eq.~\re{abw}.
For the present study the latter were fixed at the values in Eq.~\re{adef},
which give agreement with a range of data on power corrections.
In future these parameters could be varied to achieve a best fit.
For serious comparisons with data, the 2-loop modification \re{2loop}
should be used, and the varying number of active flavours would
also need to be taken into account.

A renormalon analysis of the perturbation series suggests another
way to estimate power corrections, either through Borel transformation
or by optimal truncation of the series. Although this approach seems
to involve no non-perturbative parameters, it is equivalent to
using the perturbative expression for the running coupling above a
cutoff $Q_0$ and setting it to zero below $Q_0$. Since the value of
the effective cutoff $Q_0$ depends on the details of the integrand,
viz.\ the powers of momentum and $\as$ (Table 2), this procedure
cannot correspond to any universal model for $\as$. In particular,
the effective cutoff increases rapidly with the power of $\as$, and
therefore renormalon estimates of higher-order effects will fall below
those of any universal model for sufficiently high orders.

Whether a universal model of the running coupling, including
non-perturbative terms, can be extended to higher orders is of
course open to question anyway. For the parameter values used
in Eq.~\re{adef}, the model coupling is nowhere larger
than 0.8 (Fig.~1), and so an expansion in powers of
$\as(k^2)/\pi$ does not seem unreasonable. The hope is
also that large coefficients due to infrared
renormalons should not appear in this expansion. 

\acknowledgments
I am most grateful to M.\ Dasgupta, Yu.L.\ Dokshitzer and G.\ Marchesini
for helpful conversations.

\end{document}